\documentclass[pdftex,a4paper,parskip,fleqn]{scrartcl}
\usepackage[ansinew]{inputenc}
\usepackage[T1]{fontenc}
\usepackage{graphicx}
\usepackage{struktex}
\usepackage{listings}
\usepackage{mathcomp}
\usepackage{mathtools}
\usepackage{amsmath, amsthm, amssymb}
\usepackage{epstopdf}
\usepackage{hyperref}
\usepackage{xcolor}
\usepackage[linesnumbered, ruled]{algorithm2e}
\usepackage[babel=true]{csquotes}
\newtheorem{satz}{Theorem}

\newtheorem{lem}[satz]{Lemma}
\newtheorem{defi}{Definition}
\newtheorem{rem}{Remark}

\begin{document}

\title{Finding Euler Tours in the StrSort Model}
\author{Lasse Kliemann \and Jan Schiemann \and Anand Srivastav}
\date{\small%
Department of Computer Science\\
Kiel University\\
Christian-Albrechts-Platz 4\\
24118 Kiel, Germany\\
\texttt{\{lki,jasc,asr\}@informatik.uni-kiel.de}}
\maketitle

\begin{abstract}
  \textbf{Abstract:} We present a first algorithm for finding Euler tours in undirected graphs in the StrSort model.
  This model is a relaxation of the semi streaming model.
  The graph is given as a stream of its edges and can only be read sequentially,
  but while doing a pass over the stream we are allowed to write out another stream
  which will be the input for the next pass.
  In addition, items in the stream are sorted between passes.
  This model was introduced by Aggarwal et al. in 2004.
  Here we apply this model to the problem of finding an Euler tour in a graph
  (or to decide that the graph does not admit an Euler tour).
  The algorithm works in two steps.
  In the first step, a single pass is conducted while a linear (in the number of vertices $n$) amount of RAM is required.
  In the second step, $\mathcal O(\log(n))$ passes are conducted while only $\mathcal O(\log(n))$ RAM is required.
  \par
  We use an alteration of the algorithm of Atallah and Vishkin from 1984 for finding Euler tours in parallel.
  It finds a partition of edge-disjoint circuits and arranges them in a tree expressing their connectivity.
  Then the circuits are merged according to this tree.
  In order to minimize the needed amount of RAM, 
  we evade the need to store the entire tree and use techniques suggested by Aggarwal et al.
  to exchange information concerning the merging of circuits.
\end{abstract}

\section{Introduction}

For the processing of large graphs, the \emph{graph streaming} or \emph{semi streaming} model
has been studied extensively in the last decade.
In this model, the graph is given as a stream of its edges meaning that only sequential access is possible.
Random-access memory (RAM) is restricted to $\mathcal O(n \cdot \text{polylog}(n))$ edges at a time.
This makes the model non-applicable to problems where already the size of the solution can be larger than that.
In the Euler tour problem, we are looking for a closed walk in an undirected graph such
that each edge is visited exactly once (or we wish to determine that the graph does not admit such a walk).
The solution size (in the positive case) can be of order $\Theta(n^2)$, since it contains all edges of the graph.
This problem hence calls for a relaxation of the graph streaming model.

\subsection{StrSort and W-Stream}
Aggarwal et al.~\cite{Ruhl,Aggarwal} presented a less restrictive streaming model, called \emph{StrSort-model}.
It consists of alternating streaming and sorting passes.
A streaming pass consists of a Turing machine with local memory of size $m$ and two tapes.
On one tape, the Turing machine reads a sequence $S=x_1,...,x_k$ of $k \in \mathbb N$ items.
On the other tape, an output stream is written.
On both tapes, the Turing machine can move only left-to-right.
In a sorting pass, a Turing machine with a global partial order
sorts items on a tape according to this order and gives the sorted items as output.
\begin{defi}\label{stso}
StrSort$(p_{\text{Str}} , p_{\text{Sort}} , m)$ is the class of functions computable by the composition of up to $p_{\text{Str}}$ streaming passes
and $p_{\text{Sort}}$ sorting passes, each with memory $m$, where:
\begin{itemize}
\item the local memory is maintained between streaming passes
\item streams produced at intermediate stages are of length $\mathcal O(n)$, where $n$ is the length of the input stream.
\end{itemize}
\end{defi}
Using only $\mathcal O(\text{polylog}(n))$ memory space and $\mathcal O(\text{polylog}(n))$ passes is sufficient for solving many graph problems 
in this streaming model, such as minimum spanning tree, maximal independent set and mincut~\cite{Ruhl}, hence the following definition of Aggarwal et al.:
\begin{defi}
PL-StrSort := $\cup_k \text{ StrSort }(O(\log^k n), O(\log^k n))$
\end{defi}
Demetrescu et al.~\cite{Dem09} showed for a few graph problems that the sorting steps are not necessary.
In the so-called W-Stream-model, which uses only the streaming steps (i.e. StrSort$(p_{\text{Str}} , 0, m)$), they show a tradeoff between internal memory and streaming passes for 
undirected connectivity and single-source shortest paths in directed graphs.

\subsection{Euler tours}
The Euler tour problem is one of the fundamental problems of graph theory.
Given a graph $G=(V,E)$, find an Euler tour or state that the graph is not Eulerian.
In RAM model finding Euler tours in polynomial time is relatively easy, and there are multiple well known algorithms for that task.
But the problem gets more complicated considering a big data environment in the form of a streaming or external memory model.
For the latter, an algorithm of Atallah and Vishkin \cite{Atallah} for solving Euler tours in PRAM is used.
The algorithm has a running time of $\mathcal O(\text{log}(n))$ and uses $n+m$ processors, where $n$ is the number of vertices and $m$ is the number of edges in $G$.
Since PRAM algorithms can be transferred to external memory \cite{Chiang}, 
this result can be remodeled to get an external memory algorithm solving the Euler tour problem in $\mathcal O(\text{log}(n) \text{ sort}(n+m))$ I/Os.
While the different problem \enquote{Euler tour on a tree} is regarded in multiple papers (e.g. \cite{Dem10},
also with a transfer of PRAM algorithms), 
to the best of our knowledge the classical Euler tour problem was not considered in a streaming model before.

\subsection{Our contribution}
We give the 2-step StrSort-algorithm \emph{EulerStr} for finding an Euler tour in a given graph $G=(V,E)$ with $n:=|V|$ and $m:=|E|$.
The first step is a single pass W-stream algorithm with memory space $\mathcal O(n\text{ log}(n))$,
that is, the bound which is usually used in the semi-streaming environment.
The second step is a PL-StrSort algorithm with $O(\text{log}(n))$ alternating streaming and sorting passes and $O(\text{log}(n))$ memory space.
The stream length will be $\mathcal O(m)$ the whole time.
We use the technique of Atallah and Vishkin for finding Euler tours in parallel, but with two differences:
\begin{itemize}
 \item The algorithm of Atallah and Vishkin uses memory space of a size inappropriate for a streaming environment. 
 We limit the memory space needed in the different steps using the storage of suitable subgraphs and different standard techniques of the StrSort model.
 \item In contrast to the algorithm of Atallah and Vishkin, we don't save the predecessor edge in the Euler
tour for every edge. We output the edges in the right order given by a found Euler tour. This can be
interesting for further processing the Euler tour.
\end{itemize}

\section{Preliminaries}
Let $G=(V,E)$ be an undirected graph with vertex set $V$ and edge set $E$.
A walk of length $k$ is an alternating sequence $v_1 - e_1 - v_2 - e_2 - ... - v_{k} - e_{k} - v_{k+1}$ of vertices and edges, where $e_i = \{v_i, v_{i+1}\}$
for all $i \in \{1,...,k\}$.
A trail is a walk without repeating edges, i.e. for all $i,j \in \{1,...,k\}$: $i \neq j \Leftrightarrow e_i \neq e_j$.
A circuit is a trail with the property $v_1 = v_{k+1}$, i.e. a closed trail.
An Euler tour is a circuit that uses each edge in $E$ exactly once.
A graph that contains an Euler tour is called Eulerian.
A path is a walk without repeating vertices or edges.
A cycle is a circuit with $v_i \neq v_j$ for all $i,j \in \{1,...,k\}$.

A rooted tree is a tree, in which one vertex $r$ is assigned as a root.
In a rooted tree, the depth of a vertex $v$ is the length of the unique path to its root.
The vertex $u$ adjacent to $v$, which is on the $v$-$r$-path is called predecessor of $v$.
If for an vertex $w$, $v$ is the predecessor of $w$, $w$ is called an successor of $v$.
An out-tree is a rooted, directed tree, where all edges point to the respective successor.
For an directed edge $\vec{e}=(u,v)$, $u$ is called the tail, and $v$ the head of $\vec{e}$.

For an undirected Graph $G=(V,E)$, each vertex is presented with a distinct number of the set $\{1,...,n\}$ with $n:=|V|$.
The input stream consists of the $m$ edges of $G$, given in random order.

\section{Generel idea of EulerStr}
Let $G=(V,E)$ be an undirected graph. 
Unless said otherwise, we define $n:=|V|$ and $m:=|E|$ for the rest of the paper.
The algorithm EulerStr will test, if the graph is Eulerian, and if it is, will output directed edges in order $(u_1,v_1),...,(u_m,v_m)$ with the following properties:
\begin{itemize}
 \item $x_i \in V$ for all $x \in \{u,v\}, i \in \{1,...,m\}$
 \item for all $e \in E$ there is exactly one $i \in \{1,...,m\}$ with $e = \{u_i, v_i\}$
 \item $v_i = u_{i+1}$ for all $i \in \{1,...,m-1\}$, and $v_m = u_1$
\end{itemize}
Hence the sequence $u_1 - \{u_1, v_1\} - v_1 - \{u_2, v_2\} - v_2,... - \{u_m, v_m\} - v_m$ is a closed trail that uses each edge exactly once, i.e. an Euler tour.
We will often describe walks, circuits etc. analog to this as a sequence of directed edges instead of an alternating sequence of vertices and undirected edges.
This way, when sorting edges we can sort by the label of either the head or the tail, and don't have to consider the random inner order of undirected edges. 

\begin{rem}\label{remark}
We use a slight alteration of the algorithm of Atallah and Vishkin \cite{Atallah}
It consists of three general steps:
\begin{enumerate}
 \item Partition the graph into $q$ edge-disjoint circuits $C_1,...,C_q$.
 \item Create an out-tree $T=(V',E')$ with $V' = \{w_1,...,w_q\}$ and for all $i,j \in \{1,...,q\}: (w_i, w_j) \in E' \Rightarrow C_i$ and $C_j$ share a common vertex in $G$.
 \item Iteratively: Merge all circuits presented in $T$ by vertices with odd depth with the circuit presented in $T$ by the predecessor.
\end{enumerate}
\end{rem}

Step 1 is easily done in W-stream with $\mathcal O(n \text{ log}(n))$ memory space, because $n$ Edges fit into internal memory, and every subgraph with $n$ edges contains
at least one cycle.
So iteratively, edges can be taken from the input stream until $n$ edges are present in internal memory. Then, the edges of a circuit can be found, 
written on the output stream and deleted from internal memory.
If there are edges left in internal memory after the W-stream step, the graph was not Eulerian.
Alternatively, $n$ variables can be placed in internal memory, that keep track of the degree of the vertices.
This is helpful because of the following well known result:

\begin{lem}
Let $G=(V,E)$ be an undirected graph. 
Then $G$ is Eulerian, iff every vertex has even degree and the graph is connected.
\hfill $\Box$
\end{lem}

That step 2 and 3 with additional properties are giving us an Euler tour is shown in the following lemma:
\begin{lem}\label{genlem}
Let $G = \bigcup_{1 \leq i \leq q} C_i$ be an Eulerian graph partitioned into $q$ circuits $v_{i_1} - e_i^1 -v_{i_2} - ... - v_{i_{l_i+1}} = v_{i_1}$ with $i \in \{1,...,q\}$
and $\sum_{i=1}^k l_i = m$.
$l_i$ is the length of the circuit $C_i$.
Let $T=(V',E')$ be a rooted tree with $V' = \{w_1,...,w_q\}$, root $w_1$ and for $i,j \in \{1,...,q\}$: $(w_i, w_j) \in E' \Rightarrow C_i$ and $C_j$ share a common vertex in $G$.
For every $i \in \{2,...,k\}$, let $v_{i_1}$ be a vertex that the circuit $C_i$ shares with its predecessor.
Then the following recursive algorithm gives an Euler tour:

\begin{algorithm}[ht]\caption{\label{euler-tree}Algorithm euler-tree}
S:=\{2,...,q\} (global)\;
output vertex $v_{1_1}$\;
eul-suc($C_1$)\;
\end{algorithm}

\begin{algorithm}[H]\caption{\label{eul-suc}eul-suc($C_j$)}
 $i:=1$\;
 \Repeat{$i \leq l_j$}
    {
    \If{$w_j$ has a successor $w_k$ with $k \in S$ and $v_{j_i} = v_k^1$}
       {
       $S:= S\backslash \{k\}$\;
       eul-suc($C_k$)\;
       }
    \Else
       {
       output edge $e_j^i$ and vertex $v_{j_{i+1}}$\;
       $i:=i+1$\;
       }
    }
\end{algorithm}
\end{lem}\hfill $\Box$

\begin{rem}
The route in the tree chosen by the algorithm describes an 'Euler tour on a tree' (for Definition see e.g. \cite{Mehta}).
\end{rem}

\begin{center}
\begin{picture}(0,0)%
\includegraphics{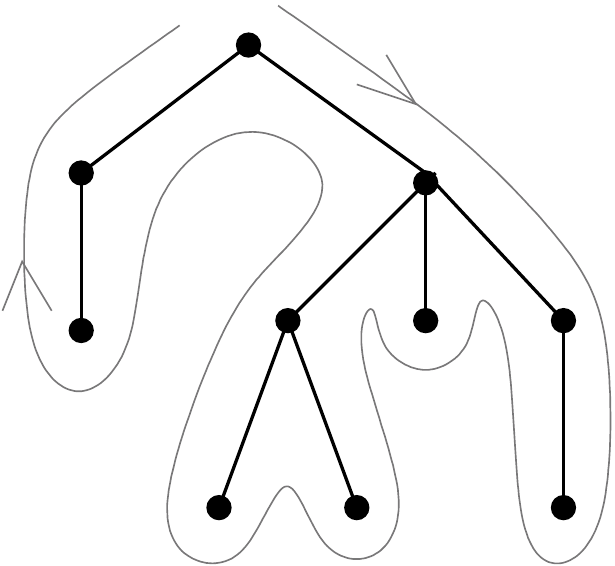}%
\end{picture}%
\setlength{\unitlength}{4144sp}%
\begingroup\makeatletter\ifx\SetFigFont\undefined%
\gdef\SetFigFont#1#2#3#4#5{%
  \reset@font\fontsize{#1}{#2pt}%
  \fontfamily{#3}\fontseries{#4}\fontshape{#5}%
  \selectfont}%
\fi\endgroup%
\begin{picture}(2804,2574)(394,-1768)
\end{picture}%

\end{center}

\textbf{Proof of lemma \ref{genlem}:}
Because of the set $S$, every vertex $w_i$ in $T$ is regarded at most once.
When $w_i$ is regarded, with EulSuc($C_i$) every edge of $C_i$ is part of the output at some point.
Now we have to show two things:
\begin{enumerate}
 \item The algorithm runs EulSuc($C_i$) for every $i \in \{1,...,q\}$.
 \item The output is an circuit of $G$.
\end{enumerate}
With both properties it is shown, that the output is an Euler tour.
We use an induction over $q$.
For $q=1$, the algorithm starts with $v_{1_1}$, and since $C_1$ is a circuit that contains all edges of $G$ in correct order, the output is an Euler tour.
Now we assume, that both properties are correct for all Eulerian graphs with partition of $q$ circuits.
Let $G$ be an Eulerian graph with partition of $q+1$ circuits. 
W.l.o.g. let $w_{q+1}$ be a leaf in the rooted tree $T$.
Then $T\backslash\{w_{q+1}\}$ is a connected graph, therefore $\tilde{G}:=G\backslash\{e_{q+1}^1,...,e_{q+1}^{l_{q+1}}\}$ is connected.
When a circuit is deleted from an Eulerian graph and the result is connected, then this graph is also Eulerian.
This graph has a partition of $q$ circuits, so by assumption the algorithm works for $\tilde{G}$.
Let $w_j$ ($j \in \{1,...,q\}$) be the predecessor of $w_{q+1}$.
Then at some point the algorithm runs EulSuc($C_j$).
Furthermore there is a $k \in \{1,...,l_j\}$ with $v_{j_k} = v_{{q+1}_1}$.
At EulSuc($C_j$) with variable $i=k$, the algorithm doesn't continue with edge $e_j^k$ until all successors of $w_j$ are taken care of.
So at some point EulSuc($C_{q+1}$) starts, proving the first property.
Since $w_{q+1}$ is a leaf, the algorithm outputs all edges of $C_{q+1}$ at once in correct order, ending again at vertex $v_{j_k}$.
Therefore, the algorithm combines an Euler tour of $\tilde{G}$ with the circuit $C_{q+1}$, resulting in an Euler tour of $G$, proving the second property.
\hfill $\Box$

Lemma \ref{genlem} shows that, if we have an Eulerian Graph, a partition into circuits $C_1,...,C_q$ and a rooted tree $T$ with the mentioned properties,
a vertex $w_i$ in $T$ can be merged with his predecessor $w_j$ by combining the circuits $C_i$ and $C_j$, i.e. inserting $C_i$ into $C_j$ at the right place.
For this, we want to make sure that the first vertex of $C_i$ is a common vertex of $C_j$, so we don't have to change the order of $C_i$ before combining it with $C_j$.
Notice that after the merging into a longer circuit $C_{j'}$, the first vertex of this circuit is still a common vertex of its predecessor,
therefore we just have to take care of the order of all circuits once.
Since in the actual algorithm EulerStr we will store circuits as a sequence of directed edges, this translates to:
The tail of the first directed edge of a circuit $C_i$ has to be the head of a directed edge of $C_j$, where $w_j$ is the predecessor of $w_i$ in $T$.

\section{The semi-W-stream step}
\subsection{Illustrating the step}
In this section, we describe the one pass step of EulerStr with $\mathcal O(n\text{ log}(n))$ memory.
In this pass, we want to perform step 1 and 2 of remark 1.
For finishing step 2, we will have to use an additional StrSort$(\mathcal O(1) , \mathcal O(1) , \text{log}(n))$-algorithm, which will be described in the
following section.

As mentioned, in the input stream we have the $m$ edges in random order.
The vertices of $G$ are called $\{v_1,...,v_n\}$.
In internal memory we keep the following variables with $\mathcal O(\text{log}(m))=\mathcal O(\text{log}(n))$ space each:
\begin{itemize}
 \item $com_i \in \{0,...,n\}$ for $i \in \{1,...,n\}$, starting with $com_i = 0$ for all $i \in \{1,...,n\}$
 \item $pre_i \in \{0,...,m\}$ for $i \in \{1,...,n\}$, starting with $pre_i = 0$ for all $i \in \{1,...,n\}$
 \item $cir \in \{0,...,m\}$, the number of circuits found yet
\end{itemize}
Additionally, we build a tree $\bar{T}$ with $\mathcal O(n)$ vertices in internal memory.
It will later be extended to the desired rooted tree $T$.

Step 1 of remark \ref{remark} is easily done as explained before.
We read up to $n$ edges, find a circuit $C$ and output the edges in correct order in relation to the circuit as well as the direction in which the respective edge is traversed.
These edges will get $4\text{ log}(m)$ additional memory space and be called \textit{'graph edges'}.
In these edges, we store the label $cir$ of the circuit the edge is in, and the position of the edge in the circuit sequence.
Occasionally, we also output \textit{'information edges'}.
The purpose and form of these information edges and the actual memory usage of the graph edges will be explained later.

For $l \in \{1,...,q\}$ let $G_l$ be the graph consisting of all vertices and edges that are used in at least one circuit $C_1,...,C_l$.
For $i \in \{1,...,n\}$, the variable $com_i$ keeps track of the connected component the vertex $v_i$ is currently in, considering the current graph $G_l$.

The variable $pre_i$ stores the label of the first circuit found that uses the vertex $v_i$.

The tree $\bar{T}$ is constructed as follows:
We create a vertex $w_l \in \bar{T}$ every time a found circle $C_l$ has at least one of the following properties:
\begin{enumerate}
 \item $pre_i = 0$ for some $i \in \{1,...,n\}$ with $v_i \in C_l$
 \item $com_i \neq com_j$ for some $i,j \in \{1,...,n\}$ with $v_i, v_j \in C_l$
\end{enumerate}
So for every circuit $C_l$ that contains a vertex not used before, or connects two connected components in the graph $G_{l-1}$, a vertex $w_l$ in $\bar{T}$ is created.
For each property, there can be at most $n$ circuits fulfilling it, so the graph $\bar{T}$ has $\mathcal O(n)$ vertices.
Edges in $\bar{T}$ are build the following way:
Let $G_i$ be the graph that contains all vertices and edges used by the circuits $C_1,...,C_i$.
If a circuit $C_{i+1}$ is found, that has vertices of the connected components $A_1,...,A_k$ in $G_i$, let $v_{i_1},...,v_{i_k} \in V$ with $v_{i_j} \in A_j$ for all $j \in \{1,...,k\}$.
Let $C_{j_1},...,C_{j_{k'}}$ be the circuits stated in $pre_{i_1},...,pre_{i_k}$, i.e. the circuits that used the vertices $v_{i_1},...,v_{i_k}$ for the first time.
Then the edges $\{w_{i+1}, w_{j_1}\}, ..., \{w_{i+1}, w_{j_{k'}}\}$ are added to $\bar{T}$.
The vertices $w_{j_1},...,w_{j_{k'}}$ exist, because the circuits $C_{j_1},...,C_{j_{k'}}$ fulfill property 1.


\textbf{Example:}

\begin{figure}[ht]
\begin{center}
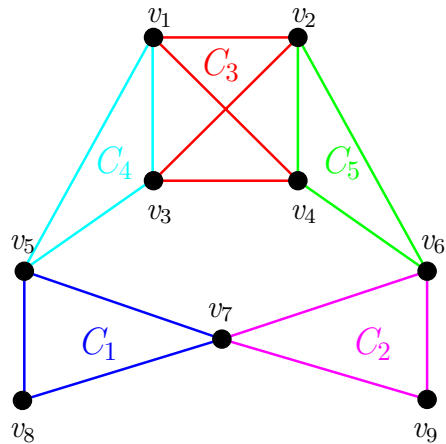
\caption{Partition into circuits (cycles here)}\label{fig1}
\label{figure:example1}
\end{center}
\end{figure}

\begin{figure}[ht]
\begin{center}
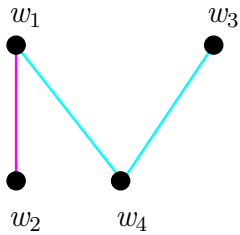
\caption{Creating the graph $\bar{T}$}\label{fig2}
\label{figure:example2}
\end{center}
\end{figure}

Figure \ref{fig1} gives an example on a graph with nine vertices $v_1,...,v_9$.
Assume that the circuits found are the cycles $C_1,...,C_5$ in that order.
$C_1$ fulfills property 1, so a vertex $w_1$ in $\bar{T}$ is created.
We set $pre_i=1$ and $com_i=1$ for $i \in \{5,7,8\}$.
$C_2$ only has property 1 and shares the vertex $v_7$ with $C_1$ (this information is stored in $pre_7$), so $com_6=1$, $com_9=1$ and
$w_2$ is created in $\bar{T}$ with edge $\{w_1,w_2\}$.
Furthermore $pre_6=2$ and $pre_9=2$, because $v_6$ and $v_9$ are used for the first time.
With $C_3$, we set $pre_i=3$ for $i \in \{1,2,3,4\}$ and have a new connected component in $G_3$ with $com_i=3$ for $i \in \{1,2,3,4\}$.
We place a vertex $w_3$ in $\bar{T}$ without additional edges.
$C_4$ only has property 2 and connects the components '1' and '3'.
Vertices $v_1$ and $v_5$ are selected with $com_5=1$ and $com_1=3$.
We create a vertex $w_4$, and since $pre_5=1$ and $pre_1=3$, we connect the vertex with edges $\{w_4,w_1\}$ and $\{w_4,w_3\}$ in $\bar{T}$.
The circuit $C_5$ has neither of the two properties, so there is no additional vertex in $\bar{T}$.
However, to get the extended graph $T$, we will store an 'information edge' in the output stream, containing the information, that 
$\bar{T}$ with vertex $w_5$ and edge $\{w_5,w_3\}$ (selected because $v_2 \in C_5$ and $pre_2 = 3$) would still be a tree.
The result is shown in figure \ref{fig2}

We have to show that the resulting graph is a tree.
In that case, the graph can be stored in internal memory

\begin{lem}
After the streaming procedure, $\bar{T}$ is a tree.
\end{lem}

\subsection{Graph edges and information edges}
We store two kinds of edges in the stream: Graph edges, which are the actual edges in $G$ with additional information, and information edges which represent the 
tree $T$.
A graph edge $e_i^k$ of circuit $C_i$ has $2\text{ log}(n) + 4\text{ log}(m)$ memory space and is at first set up as follows ($l_i$ is the length of circuit $C_i$):
\begin{align}\label{graphedge}
 e_i^k:=(v_{i_k}, v_{i_{k+1}}, i, k, 0, 0) \text{ for } k \in \{1,...,l_i\}
\end{align}
\begin{itemize}
 \item $\{v_{i_k}, v_{i_{k+1}}\}$ is the original edge in $G$.
 \item $e_i^k \in C_{i}$ and walking on $C_{i}$, $e_i^k$ is passed from $v_{i_k}$ to $v_{i_{k+1}}$.
 \item $k$ is the placement of $e_i^k$ in $C_{i}$ in the order stored in the output stream.
 \item Later when merging circuits, the last two variables will help representing the predecessor circuit $C_j$ and the placement $k'$ of the edge of $C_j$, behind
 which the circuit $C_i$ will be inserted.
\end{itemize}

Information edges are the edges build in $\bar{T}$ and later $T$. 
They also contain additional information.
Since we need a rooted tree, variables concerning this are placed in these edges.
They have $\text{ log}(n) + 4\text{ log}(m)$ memory space and are build as follows:
\begin{align}\label{infoedge}
f_i^j:=(i, j, d_i, v ,p_{i})
\end{align}
\begin{itemize}
 \item $f_i^j$ represents the edge $\{w_{i}, w_{j}\} \in T$ and $w_{i}$ is the predecessor of $w_{j}$ in $T$.
 \item $d_{i}$ is the depth of $w_{i}$ in $T$.
 \item $v$ is a common vertex of $C_{i}$ and $C_{j}$ in $G$.
 \item Similar to graph edges, $p_{i}$ will be the placement of the edge in $C_{i}$, which has $v$ as its head, so when merging $C_{i}$ and $C_{j}$, 
 this can be done by inserting $C_{j}$ into $C_{i}$ behind this edge.
 But for now, this memory space will be used for storing different variables.
\end{itemize}

\subsection{The algorithm}

\begin{algorithm}[H]\caption{\label{circuit-find}Algorithm circuit-find}
  \SetKwData{Left}{left}
  \SetKwData{Up}{up}
  \SetKwFunction{FindCompress}{FindCompress}
  \SetKwInOut{Input}{input}
  \SetKwInOut{Output}{output}
  
\Indm
\Input{Undirected graph $G=(\{v_1,...,v_n\},E)$ with edges in random order, $m:=|E|$}
\Output{$m$ graph edges and $q$ information edges for $q \leq m$}
\Indp
  \BlankLine
 $com_i := 0$ for all $i \in \{1,...,n\}$\;
 $pre_i := 0$ for all $i \in \{1,...,n\}$\;
 $cir:=0$\;
 $s:=false$, $s_{cr}:=false$ \tcp*{indicates if vertex in $\bar{T}$ will be or is created}
 $s_{edge}:=0$ \tcp*{indicated potential edge in $\bar{T}$}
 $s_{vert}:=0$ \tcp*{indicated common vertex in $G$}
 $\bar{T} := (\bar{V'}, \bar{E'})$, $\bar{T}:=\emptyset, \bar{T}:=\emptyset$\;
 $S_{comp} := \{ 0 \}$ \tcp*{keeps track of conn\text{.} comp\text{.} concerning current circuit}
 $com^* := 0$\;
 \Repeat{(end of stream) and (no edges in internal memory)}
 {
    read stream until ($n$ edges are in internal memory) or (end of stream)\;
    find circuit $C=v_{i_1}-e_i^1-...-v_{i_{l_i}}-e_i^{l_i}-v_{i_1}$ with vertices $v_{{i'}_1},...,v_{{i'}_{l'}}$ ($l_i, l' \in \mathbb N$) \;
    if there is no such circuit, \Return{'graph is not Eulerian'}\;
    $cir:=cir+1$\;
    \mbox{\textsc{new-test}}($C$)\;
    \mbox{\textsc{comp-test}}($C$)\;
    \If{s=false}{output information edge $(s_{edge}, cir, 0, v_{s_{vert}} ,1)$\;
    sort $C$, s.t. $C=v_{i_1}-e_1-...-v_{i_{l_i}}-e_{l_i}-v_{i_1}$ with $v_{i_1}=v_{s_{vert}}$\;
    }
        \For{j:=1 to $l_i$-1}{output graph edge $(v_{i_j}, v_{i_{j+1}}, cir, j, 0, 0)$\;}
         output graph edge $(v_{i_{l_i}}, v_{i_1}, cir, l_i, 0, 0)$\;
  delete $C$ in internal memory\;  
  $s:=false$, $s_{cr}:=false$, $s_{edge}:=0$, $s_{vert}:=0$, $S_{comp} := \{ 0 \}$, $com^* := 0$\;
 }
 \For{i:=1 to n-1}
 {\If{$com_i \neq com_{i+1}$}{\Return{'graph is not Eulerian'}} }
 write $\bar{T}$ as rooted tree\;
 for every $w_i \in \bar{V'}$, let $d_i$ be the depth of $w_i$ in $\bar{T}$\;
 for every information edge $(i, j, 0, v, 0)$ in internal memory output information edge $(i, j, d_i, v, 0)$\;
\end{algorithm}

\begin{algorithm}\caption{\label{new-test}Algorithm new-test}
 \For{j:=1 to $l'$}
    {\If{$pre_{i'_j} = 0$}
       {
       $s:=true$\;
       $pre_{i'_j}:=cir$\;
       }
      \Else{\If{$s_{edge} = 0$}{$s_{edge}:=pre_{i'_j}$\; $s_{vert}:=i'_j$\; $S_{comp}:=S_{comp} \cup \{com_{i'_j}\}$\; $com^* := com_{i'_j}$} }
    }
    \If{$s=true$}
      {create vertex $w_{cir}$, $\bar{V'} :=\bar{V'} \cup \{w_{cir}\}$\;
       \If{$s_{edge} \neq 0$}{create edge $\{w_{cir}, w_{s_{edge}}\}$, $\bar{E'} :=\bar{E'} \cup \{ \{w_{cir}, w_{s_{edge}}\} \}$\;
        create information edge $(s_{edge}, cir, 0, v_{s_{vert}} ,0)$\;
       }
       \Else{
       \For{j:=1 to $l'$}{$com_{i'_j} := cir$}
       }
      }
\end{algorithm}

\begin{algorithm}\caption{\label{comp-test}Algorithm comp-test}
 \If{$com^* \neq 0$}
       {
          \For{j:=1 to l'}
          {
          \If{$com_{i'_j} \neq com^*$}
            {
          
             \If{$s=false$}
               {
                $s:=true$\;
                create vertex $w_{cir}$, $\bar{V'} :=\bar{V'} \cup \{w_{cir}\}$\;
                create edge $\{w_{cir}, w_{s_{edge}}\}$, $\bar{E'} :=\bar{E'} \cup \{ \{w_{cir}, w_{s_{edge}}\} \}$\;
                create information edge $(s_{edge}, cir, 0, v_{s_{vert}} ,0)$\;
                }
              \If{$com_{i'j} \notin S_{comp}$}
              {
               create edge $\{w_{cir}, w_{pre_{i'_j}}\}$, $\bar{E'} :=\bar{E'} \cup \{ \{w_{cir}, w_{pre_{i'_j}}\} \}$\;
               create information edge $(pre_{i'_j}, cir, 0, v_{{i'}_j} ,0)$\;
               $S_{comp}:=S_{comp} \cup com_{i'_j}$\;
              }
            }
          
          }
       \For{k:=1 to n}
       {
       \If{$com_k \in S_{comp}\backslash\{com^*\}$ }
          {
          $com_k:=com^*$\;
          }
       }
       }
\end{algorithm}

\begin{rem}
When algorithm circuit-find found a circuit $C_i$ in step 12, it is tested if $C_i$ uses a vertex of $G$ for the first time (new-test) 
or connects connected components in $G_{i-1}$ (comp-test).
In new-test, step 2 to 5 test if a vertex is used for the first time.
If this is the case, $s$ indicates that a new vertex $w_i$ is created in the tree $\bar{T}$.
Step 6 to 13 test if the circuit uses a vertex used by a circuit $C_j$ before.
If $w_i$ is created, an edge $\{w_i, w_j\}$ is stored and an information edge is output (step 17 to 20).
$S_{comp}$ keeps track of the connected components in $G_{i-1}$ touched by $C_i$.
If $C_i$ only uses new vertices, they will be a connected component in $G_i$.
This is noted in step 21 to 25.
Algorithm comp-test starts if $C_i$ uses a vertex used before.
Let $A_k$ be the connected component of that vertex in $G_{i-1}$.
In comp-test it is tested if $C_i$ uses vertices, which are not in $A_k$ and not used for the first time.
If this happens for the first time, and there is not already a vertex $w_i$ in $\bar{T}$, 
such a vertex is created in step 4 to 9 with the necessary graph and information edge.
Otherwise, just the graph and information edge is made.
In step 17 to 21 the variables $com_k$ are updated.
If after new-test and comp-test there is still no vertex $w_i$ in $\bar{T}$, in step 17 to 20 of circuit-find an information edge is output.
The last entry is '1', indicating that $C_i$ has no representative in $\bar{T}$.
In step 19, the circuit is output such that the tail of the first edge is a common vertex of the circuit noted in the information edge.
The connectivity of $G$ is tested in step 28 to 32.
Finally the rooted tree is build, and the stored information edges are updated and output.
\end{rem}

\section{PL-StrSort algorithm}
\subsection{Merging circuits}\label{expl-merg}
The information edges indicate a rooted tree $T$ like in lemma \ref{genlem}.
Let us have two circuits $C_i$, $C_j$ and an information edge $e=(i, j, d, v, p)$, where $w_i$ is the predecessor of $w_j$ in $T$, 
$d$ is the depth of $w_i$ in $T$, $v \in V$ is a common vertex of $C_i$ and $C_j$ in $G$ and $p \in \mathbb N$ is the position of an edge in $C_i$ 
which has $v$ as its head.
If $v$ is the tail of the first edge representing $C_j$, then the two circuits can be merged in the following way:

The graph edges of $C_i$ stay the same with $e_i^k:=(v_{i_k}, v_{i_{k+1}}, C_{i}, k, 0, 0)$ for $k \in \{1,...,l_i\}$ and the length of the circuit $l_k$, and
the graph edges of $C_j$ are changed to $e_j^k:=(v_{j_k}, v_{j_{k+1}}, C_{i}, p, c_{j}, k)$ for $k \in \{1,...,l_j\}$.
When sorting these edges by the size of the four last labels (from left to right), both circuits are placed in the same region because of the label $C_i$.
Furthermore with label 4, $C_j$ is placed between the edges $p$ and $p+1$ of $C_i$, and since edge $p$ of $C_i$ has the common vertex $v$ as its head 
and $v_{j_1} = v$, the resulting order is a circuit containing the edges of $C_i$ and $C_j$.
With the labels 5 and 6, inner order of $C_j$ is maintained, even if multiple circuits are inserted at position $p$ of circuit $C_i$.

Getting the informations needed for the graph edges of $C_j$ to be changed is the task of the information edge.
But first we have to take care of a few things that couldn't be finished in the last algorithm.

\subsection{Preparations}\label{prep}
We are missing a few key points for the merging to work:

\begin{enumerate}
 \item Every circuit $C_i$ with $w_i \in \bar{T}$ was output before the predecessor in $T$ was decided. 
 The orders of their graph edges have to be changed, so that the tail of the first edge is a common vertex with the predecessor in $T$.
 \item The information edges with a vertex not contained in $\bar{T}$ were output before the rooted tree was made, so they miss the information 
 about the depth of the predecessor in $T$.
 \item All information edges lack the last information: The position of the graph edge of the predecessor circuit, behind which the successor circuit will be inserted.
\end{enumerate}

3. won't be a problem.
The algorithm will iteratively merge circuits and produce information edges belonging to a rooted tree $T'$ with height about half the height of the original tree $T$.
At that point, the information edges will again miss the information about graph edge positions.

We will now show StrSort algorithms with respectively $\mathcal O(1)$ passes and $\mathcal O(\text{log}(n))$ memory space for each of problem 1 and 2.
Analog to the strategies in \cite{Ruhl} and \cite{Aggarwal}, 
the sorting step is used to put edges needing information next to edges having said information, 
so both can be put in internal memory for information transfer during the next streaming step.

\subsubsection{Rotating circuits}
Let $C_j$ be a circuit with $w_j \in \bar{T}$.
If $d_j > 0$, $w_j$ has a predecessor $w_i$ in $\bar{T}$.
The information edge $f_i^j$ contains a common vertex $v$ of $C_i$ and $C_j$, 
but the order of $C_j$ stored in the graph edges wasn't changed according to $v$ during algorithm circuit-find.
The order of $C_j$ can be changed as follows:
\begin{itemize}
 \item Sort the graph edges by circuit label and placement, and the information edges by successor circuit 
 s.t. in the stream a circuit is stored directly behind the information edge with the regarding successor circuit.
 \item While streaming a circuit $C_j$, keep the information edge $f_i^j$ and the first graph edge $e_j^1$ of the circuit in internal memory.
 Count the number $l_j$ of edges in the circuit, and find the placement $p$ of the edge with $v$ as its tail.
 Store both informations in the last two entries of $e_j^1$.
 \item Output and delete $f_i^j$ and $e_j^1$ after reaching the next circuit in the stream (in most cases an information edge).
 Continue with the next circuit.
 \item Sort the same way as before.
 \item The necessary informations $l_j$ and $p$ are stored in $e_j^1$. 
 In the next streaming step, after reaching $e_j^1$ and storing these informations, 
 output $(v_{j_1}, v_{j_2}, j, ((k-p)\text{ mod }l_j ) + 1, 0, 0)$ and delete $e_j^1$.
 \item Read graph edges $e_j^k:=(v_{j_k}, v_{j_{k+1}}, j, k, 0, 0)$
 and output $(v_{j_k}, v_{j_{k+1}}, j, ((k-p)\text{ mod }l_j ) + 1, 0, 0)$ for $k \in \{2,...,l_j\}$.
 \item Delete $p$ and $l_j$. Continue with the next circuit. 
\end{itemize}

\subsubsection{Information edges and depth}
Let $C_j$ be a circuit with $w_j \notin \bar{T}$.
Then there is exactly one information edge with second entry $j$.
Let $C_i$ be the stored predecessor circuit and $f_i^j$ be the concerning information edge.
Then $w_i \in \bar{T}$, and the last entry of $f_i^j$ is '$1$'.
There are two cases:
\begin{itemize}
 \item $w_i$ is the root of $\bar{T}$. Then $d_j = 0$.
 \item $w_i$ has a predecessor $w_k$ in $\bar{T}$. Then the information edge concerning $\{w_i, w_k\}$ contains the depth $d_k$ of $w_k$.
 It is $d_j = d_k + 1$.
\end{itemize}
With two simple stream steps and one sort step $f_i^j=(i,j,0,v,1)$ for some $v \in V$ will get the needed information from $f_k^i$ if existing:

\begin{itemize}
 \item Change $f_i^j=(i,j,0,v,1)$ to $(j,i,0,v,1)$, i.e. change predecessor and successor, and mark that at the last variable of $f_i^j$.
 \item Sort the information edges lexicographically according to the successor (the second entry) and the last entry.
 \item The information edges with second entry '$i$' will now appear consecutively on the next input stream.
 \item If before $(j,i,0,v,1)$, there is no edge with a '$0$' as last entry and second entry '$i$', output a depth of $0$, i.e. $(i,j,0,v,0)$
 \item If there is an edge with a '$0$' as last entry, e.g. $(k,i,d_k,v,0)$, 
       then for all edges $(j,i,0,v,1)$ with $i$ as second entry output $(i,j,d_k+1,v,0)$
\end{itemize}

\subsection{The merging step}
Now we come to the merging step explained in section \ref{expl-merg}.
Due to algorithm circuit-find and the two preparation steps, the graph edges and information edges have the following properties:
\begin{enumerate}
 \item For the $q$ circuits found, let $i \in \{1,...,q\}$.
 Then circuit $C_i$ of length $l_i$ is represented by the $l_i$ graph edges $e_i^j = (v_{i_j}, v_{i_{j+1}}, i, j, 0, 0)$ for $j \in \{1,...,l_i-1\}$
 and $e_i^{l_i} = (v_{i_{l_i}}, v_{i_1}, i, l_i, 0, 0)$.
 \item Let $T=(V', \vec{E'})$ with $V':=\{w_1,...,w_q\}$ and $(\vec{e_i^j} \in \vec{E'} \Leftrightarrow$ there exists an information edge with circuit entries $i$ and $j$ in that order).
Then $T$ is an out-tree on $q$ vertices. Let $h$ be the height of $T$.
 \item For $i,j \in \{1,...,q\}$ let $f_i^j$ be an information edge.
 Then the edge has the form $f_i^j = (i, j, d_i, v, 0)$, where $w_i$ is the predecessor of $w_j$ in $T$, $d_i$ is the depth of $w_i$ and $v$ is a common vertex of $C_i$ and $C_j$.
 Furthermore $v_{j_1} = v$.
 \end{enumerate}

The algorithm will output graph edges and information edges s.t. these properties are still fulfilled 
and the out-tree represented by the information edges has height $\lfloor h/2 \rfloor$.
The number of graph edges will stay the same, still representing the edges of $G$.
After $\mathcal O(\text{log}(h)) = \mathcal O(\text{log}(n))$ iterations of the algorithm, the underlying out-tree has a height of 0, 
so the graph edges form a single circuit i.e. an Euler-tour of $G$.\\

\begin{algorithm}[H]\caption{\label{tree-merge}Algorithm tree-merge}
  \SetKwData{Left}{left}
  \SetKwData{Up}{up}
  \SetKwFunction{FindCompress}{FindCompress}
  \SetKwInOut{Input}{input}
  \SetKwInOut{Output}{output}
  
\Indm
\Input{Graph edges $e_i^j$ for some $i,j \in \{1,...,n\}$ and information edges $f_i^j$ for some $i,j \in \{1,...,m\}$ fulfilling the properties above with a graph $T$ of height $h$}
\Output{Graph edges and information edges representing an out-tree $T'$ of height $\lfloor h/2 \rfloor$ and fulfilling the properties above}
\Indp
  \BlankLine
$count:=0$\;
\For{all $f_i^j=(i,j,d_i,v,0)$ with $d_i$ odd}{change information edge to $(j,i,d_i,v,1)$\;}

sort:\linebreak
- information edges in front of graph edges\linebreak
- information edges: $(i_1,j_1,d_{i_1}, v_1, x_1) < (i_2,j_2,d_{i_2}, v_2, x_2) \Leftrightarrow (j_1 < j_2)$ or $(j_1=j_2 \text{ and } x_1 < x_2)$
or $(j_1=j_2 \text{ and } x_1 = x_2 \text{ and } i_1 < i_2)$\linebreak
- order of graph edges does not matter\\

stream:
\For{every information edge $(i,j,d_i,v,0)$ (with 0 as last entry)}
{
store $i$ in internal memory and output $(i,j,d_i,v,0)$ \;
as long as information edges of form $(i',j, d_j, v', 1)$ are read, output $(i, i', d_j, v', 0)$ instead\;
}

sort:\linebreak
- information edges with odd depth after every other edge, order does not matter\linebreak
- information edges, even depth: $(i_1,j_1,d_{i_1}, v, 0) < (i_2,j_2,d_{i_2}, v', 0) \Leftrightarrow (i_1 < i_2)$ or $((i_1 = i_2) \text{ and } (v < v'))$\linebreak
- graph edge and information edge with even depth:
      $(v_{i_j}, v_{i_{(j+1)}}, i, j, 0, 0) < (i',j',d_{i'}, v', 0) \Leftrightarrow (i < i')$ or $((i = i') \text{ and } (v_{i_{(j+1)}} \leq v'))$\linebreak
- graph edges: $(v_{i_j}, v_{i_{(j+1)}}, i, j, 0, 0) < (v_{i'_{j'}}, v_{i'_{(j'+1)}}, i', j', 0, 0) \Leftrightarrow 
 ((i < i')$ or $(i = i') \text{ and } (v_{i_{(j+1)}} < v_{i'_{(j'+1)}}))$ or $(i = i') \text{ and } (v_{i_{(j+1)}} < v_{i'_{(j'+1)}}) \text{ and } (j < j'))$\\

stream: \For{every graph edge $(v_{i_j}, v_{i_{j+1}}, i, j, 0, 0)$}
{
read all information edges of even depth until the next graph edges follows\;
for each such information edge $(i',j',d_{i'}, v', 0)$, output $(i',j',d_{i'}, v', j)$ instead\;
}

sort:\linebreak
- graph edges: $(v_{i_j}, v_{i_{j+1}}, i, j, 0, 0) < (v_{{i'}_{j'}}, v_{{i'}_{j'+1}}, i', j', 0, 0) \Leftrightarrow (i<i')$ or $(i=i'\text{ and }j<j')$\linebreak
- information edges: $(i_1,j_1,d_{i_1}, v, x_1) < (i_2,j_2,d_{i_2}, v', x_2) \Leftrightarrow (j_1 < j_2)$\linebreak
- information edge and graph edge: $(i',j',d_{i'}, v', 0) < (v_{i_j}, v_{i_{j+1}}, i, j, 0, 0) \Leftrightarrow (j' \leq j)$\\

stream: \For{every information edge $(i',j',d_{i'},v',x)$ with even $d_{i'}$}
{
store $i'$ and $x$ in internal memory, delete the information edge \textit{without} output\;
as long as graph edges $(v_{i_j}, v_{i_{j+1}}, i, j, 0, 0)$ are read, output $(v_{i_j}, v_{i_{j+1}}, i', x, i, j)$ instead\;
}
\mbox{\textsc{tree-merge2}}\;    
\end{algorithm}

\begin{algorithm}\caption{\label{tree-merge2}Continuation tree-merge2}
sort:\linebreak
- information edges in front of graph edges, order does not matter\linebreak
- graph edges: $(v_{i_j}, v_{i_{j+1}}, \bar{i}, \bar{j}, i, j) < (v_{{i'}_{j'}}, v_{{i'}_{j'+1}}, \bar{i'}, \bar{j'}, i', j') \Leftrightarrow
(\bar{i} < \bar{i'})$ or $(\bar{i} = \bar{i'} \text{ and } \bar{j} < \bar{j'})$ or
$(\bar{i} = \bar{i'} \text{ and } \bar{j} = \bar{j'} \text{ and } i<i')$ or
$(\bar{i} = \bar{i'} \text{ and } \bar{j} = \bar{j'} \text{ and } i=i' \text{ and } j<j')$\\
stream:\\
\For{every information edge $(i,j,d_i,v,0)$}{change to $(i,j,((d_i-1)/2),v,0)$\;}
\Repeat{end of stream}
{
   $count:=2$\;
   read graph edge $(v_{i_j}, v_{i_{j+1}}, i, j, 0, 0)$\;
   store $i$, output $(v_{i_j}, v_{i_{j+1}}, i, 1, 0, 0)$ and delete graph edge.
   \Repeat{graph edge is read that doesn't have $i$ as entry 3}
   {
   read graph edge $(v_{{i'}_{j'}}, v_{{i'}_{j''}}, i, x, i', y)$, output $(v_{{i'}_{j'}}, v_{{i'}_{j''}}, i, count, 0, 0)$ and delete graph edge\;
   $count:=count+1$\;
   }
}
\end{algorithm}

\begin{rem}
Since we  merge circuits $C_i$ with its predecessor circuit, iff $d_i$ is odd, the information edges with odd predecessor depth are not used in this iteration.
Instead, these edges have to be prepared for the next iteration.
Steps 5 to 9 are for that purpose.
Information edges with odd predecessor depth store the predecessor of the predecessor, because that will be the predecessor in the next iteration.
In step 10 to 14, the information edges concerning circuit merges get to know the placement in which the successor circuit has to be inserted.
The information edges share this knowledge with the graph edges in step 15 to 19.
In tree-merge2, the circuit insertions take place, and the graph edges are renamed according to their new circuit and placement.
\end{rem}

\begin{lem}
Including the preparation algorithms of section \ref{prep}, algorithm tree-merge is a 
PL-StrSort algorithm with $\mathcal O(\text{log}(n))$ alternating streaming and sorting passes and $\mathcal O(\text{log}(n))$ memory space.
\end{lem}

\begin{satz}
Algorithm EulerStr, consisting of 'circuit-find', preparation steps and 'tree-merge' has the following properties:
\begin{enumerate}
 \item In an undirected graph it finds an Euler-tour, if existing.
 \item The first part is a single step W-stream algorithm with $\mathcal O(n \text{ log}(n))$ memory space.
 \item The second part is a PL-StrSort algorithm.
 \item The stream never exceeds a length of $\mathcal O(m)$.
\end{enumerate}
\end{satz}

\section{Conclusion}
We have presented an algorithm for finding Euler tours in undirected graphs in the StrSort model.
It uses a single pass preparation step with $\mathcal O(n \text{log}(n))$ memory space, followed by a PL-StrSort algorithm.
With this result, various open questions appear:
\begin{itemize}
 \item Can the preparation step be replaced by an StrSort algorithm using $\mathcal O(\text{log}(n))$ passes and memory space?
 In this case, the Euler tours problem could be solved entirely by a PL-StrSort algorithm.
 However, as implied by Ruhl (\cite{Ruhl}) finding cycles might be difficult.
 \item Are there more problems where a single pass with larger memory enables it to be solved by a PL-StrSort algorithm?
 Such a preparation step might be a useful addition to the StrSort model.
 \item Since the algorithm of Atallah and Vishkin can be used for directed graphs, can our algorithm be altered to work on them?
 A direct transfer is not possible, because we can't find directed cycles in one pass with only $\mathcal O(n \text{ log}(n))$ memory space.
 We need to look for possibilities for finding directed cycles in the StrSort model.
 \item With the algorithm of Atallah and Vishkin an external memory algorithm can be designed which uses 
 $\mathcal O (\text{log}(n) \text{ sort}(n+m))$ I/O steps for finding an Euler tour.
 Since for $\mathcal O(\text{log}(n))$ memory space the StrSort model is more restrictive than the external memory model, can our technique be transferred to external memory to improve the current result?
 Again for this we have to run the preparation step with less memory space and probably more passes.
\end{itemize}

\end{document}